\documentclass[aps,prl,reprint,twocolumn,groupedaddress,superscriptaddress,nofootinbib]{revtex4-1}

\usepackage{graphics}
\usepackage[mathcal]{euscript}
\usepackage[latin1]{inputenc}
\usepackage{latexsym}
\usepackage{amsmath}
\usepackage{hyperref}   
\usepackage{graphicx}   
\usepackage{verbatim}  
\usepackage{color}      
\usepackage{subfigure}  
\usepackage{hyperref}   
\usepackage{bm}
\usepackage{graphicx}
\usepackage{amssymb}
\usepackage{bbm}
\usepackage{float}
\usepackage{slashed}
\usepackage{amsfonts}
\usepackage{euscript}
\usepackage{mathrsfs} 
\usepackage{bbding}
\usepackage{pifont}
\usepackage{cancel}
\usepackage{amssymb,fge}
\usepackage{pifont}

\usepackage[usenames,dvipsnames]{xcolor}
\hypersetup{colorlinks=true, citecolor=MidnightBlue, linkcolor=MidnightBlue,
urlcolor=MidnightBlue}

\begin{document}

\title{Opening the Pandora's box at the core of black holes}

\author{Ra\'ul Carballo-Rubio}
\email{raul.carballorubio@sissa.it}
\affiliation{SISSA - International School for Advanced Studies, Via Bonomea 265, 34136 Trieste, Italy}
\affiliation{IFPU - Institute for Fundamental Physics of the Universe, Via Beirut 2, 34014 Trieste, Italy}
\affiliation{INFN Sezione di Trieste, Via Valerio 2, 34127 Trieste, Italy}
\author{Francesco Di Filippo}
\email{francesco.difilippo@sissa.it}
\affiliation{SISSA - International School for Advanced Studies, Via Bonomea 265, 34136 Trieste, Italy}
\affiliation{IFPU - Institute for Fundamental Physics of the Universe, Via Beirut 2, 34014 Trieste, Italy}
\affiliation{INFN Sezione di Trieste, Via Valerio 2, 34127 Trieste, Italy}
\author{Stefano Liberati}
\email{liberati@sissa.it}
\affiliation{SISSA - International School for Advanced Studies, Via Bonomea 265, 34136 Trieste, Italy}
\affiliation{IFPU - Institute for Fundamental Physics of the Universe, Via Beirut 2, 34014 Trieste, Italy}
\affiliation{INFN Sezione di Trieste, Via Valerio 2, 34127 Trieste, Italy}
\author{Matt Visser}
\email{matt.visser@sms.vuw.ac.nz}
\affiliation{School of Mathematics and Statistics, Victoria University of Wellington; PO Box 600, Wellington 6140, New Zealand}


\bigskip
\begin{abstract}
\noindent
Unless the reality of spacetime singularities is assumed, astrophysical black holes cannot be identical to their mathematical counterparts obtained as solutions of the Einstein field equations. Mechanisms for singularity regularization would spark deviations with respect to the predictions of general relativity, although these deviations are generally presumed to be negligible for all practical purposes. Nonetheless, the strength and nature of these deviations remain open questions, given the present uncertainties about the dynamics of quantum gravity. We present here a geometric classification of all spherically symmetric spacetimes that could result from singularity regularization, using a kinematic construction that is both exhaustive and oblivious to the dynamics of the fields involved. Due to the minimal geometric assumptions behind it, this classification encompasses virtually all modified gravity theories, and any theory of quantum gravity in which an effective description in terms of an effective metric is available. The first noteworthy conclusion of our analysis is that the number of independent classes of geometries that can be constructed is remarkably limited, with no more than a handful of qualitatively different possibilities. But our most surprising result is that this catalogue of possibilities clearly demonstrates that the degree of internal consistency and the strength of deviations with respect to general relativity are strongly, and positively, correlated. Hence, either quantum fluctuations of spacetime come to the rescue and solve these internal consistency issues, or singularity regularization will percolate to macroscopic (near-horizon) scales, radically changing our understanding of black holes and opening new opportunities to test quantum gravity.

\end{abstract}

\maketitle

\noindent
Black holes are nowadays celebrated members of the club of compact astronomical objects. Long gone are the times when the idiosyncrasies of these solutions of the Einstein field equations cast doubts about their physical existence. Indeed, it seems fair to say that there has been a complete shift in the way that these idiosyncrasies are perceived, in particular regarding the accompanying spacetime singularities. Whenever the theoretical concept of a black hole is invoked in order to explain astronomical observations, for instance of the center of our own galaxy, the corresponding holes in the fabric of spacetime that general relativity predicts never raise eyebrows. There is a good mathematical reason for this: all indications show that general relativity does an excellent job of hiding these singularities, as the cosmic censorship hypothesis makes more precise~\cite{Penrose1999}. Moreover, these singularities were never physically expected to be there to start with, as accepting such a singular behaviour seems abhorrent from a physical standpoint. The existence of a mechanism that operates at the fringes of general relativity and rectifies the singular tendencies of the latter, is often (and, in many cases, implicitly) assumed. Whatever price one must pay for this, it may seem to be a small token compared to the issues that the acceptance of singularities would entail.

Actually, there is no question that singularity regularization will trigger observable differences with respect to the predictions of general relativity; the real question is how observationally important these differences will be. As stated above, the standard approach to this issue consists of assuming that these differences will be extremely small and, in particular, irrelevant for most astrophysical purposes. However, the truth is that this really is uncharted territory, with no rigorous theorems to guide our intuition, and with only a few glimpses of what a consistent picture may look like that have been extracted from candidate theories of quantum gravity. In this Letter, we communicate the outcome of a systematic analysis based on a well-known result in the framework of general relativity, the Penrose incompleteness theorem~\cite{Penrose1964}; we use it as guidance in order to shed light on these questions, by classifying all spacetime geometries that can arise from singularity regularization. One remarkable consequence of our analysis is that there is no room for complacency regarding singularity regularization: All the geometries that result in small deviations with respect to general relativity are internally inconsistent; these inconsistencies can be ameliorated only by accepting large deviations that should have definite observational implications. Singularity regularization cannot be invoked just to hide the ailments of general relativity under the rug; instead it comes with important physical consequences that must be dealt with.

\noindent
\textsl{Setting the stage.--}We will be dealing with 4-dimensional spacetimes $\mathscr{M}$. For simplicity, we restrict our discussion to spherical symmetry, though a number of aspects of the formalism developed below are naturally adaptable to more general situations. The isometry group of spherically symmetric spacetimes permits one to identify a foliation by 2-spheres $\mathbb{S}^2$.

\noindent
Aside from these basic requirements, we assume the following conditions as the basis of our analysis:
\begin{itemize}
\itemsep-3pt
\item[(1)]{Pseudo-Riemannian geometry provides an effective description of spacetime.}
\item[(2)]{The spacetime is globally hyperbolic, with a non-compact Cauchy surface $\mathscr{C}^3$.}
\item[(3)]{The spacetime is geodesically complete.}
\item[(4)]{There are no curvature singularities.}
\end{itemize}
The first assumption is self-explanatory, and provides the mathematical framework for our discussion. In physical terms, it implies that quantum fluctuations of the spacetime geometry must remain sufficiently small throughout dynamical evolution; this has been shown to be the case for large sectors of initial conditions in frameworks such as loop quantum gravity and cosmology (e.g.,~\cite{Kaminski2010,Ashtekar2011,Ashtekar2015}). Nevertheless, we advise the reader to keep in mind that this may not be the case~\cite{Ashtekar2005,Perez2014}, and we will discuss the possible implications towards the end of this Letter. 

The second assumption is the standard characterization of a well-posed initial-value problem, regardless of its specific dynamical details, which are left unconstrained (in particular, we are not assuming the Einstein field equations, or any constraint on the nature and behavior of matter fields). 

The third and fourth assumptions are just our specific formalization of these spacetimes not being singular.

\begin{figure}[!h]%
\begin{center}
\vbox{\includegraphics[width=0.5\columnwidth]{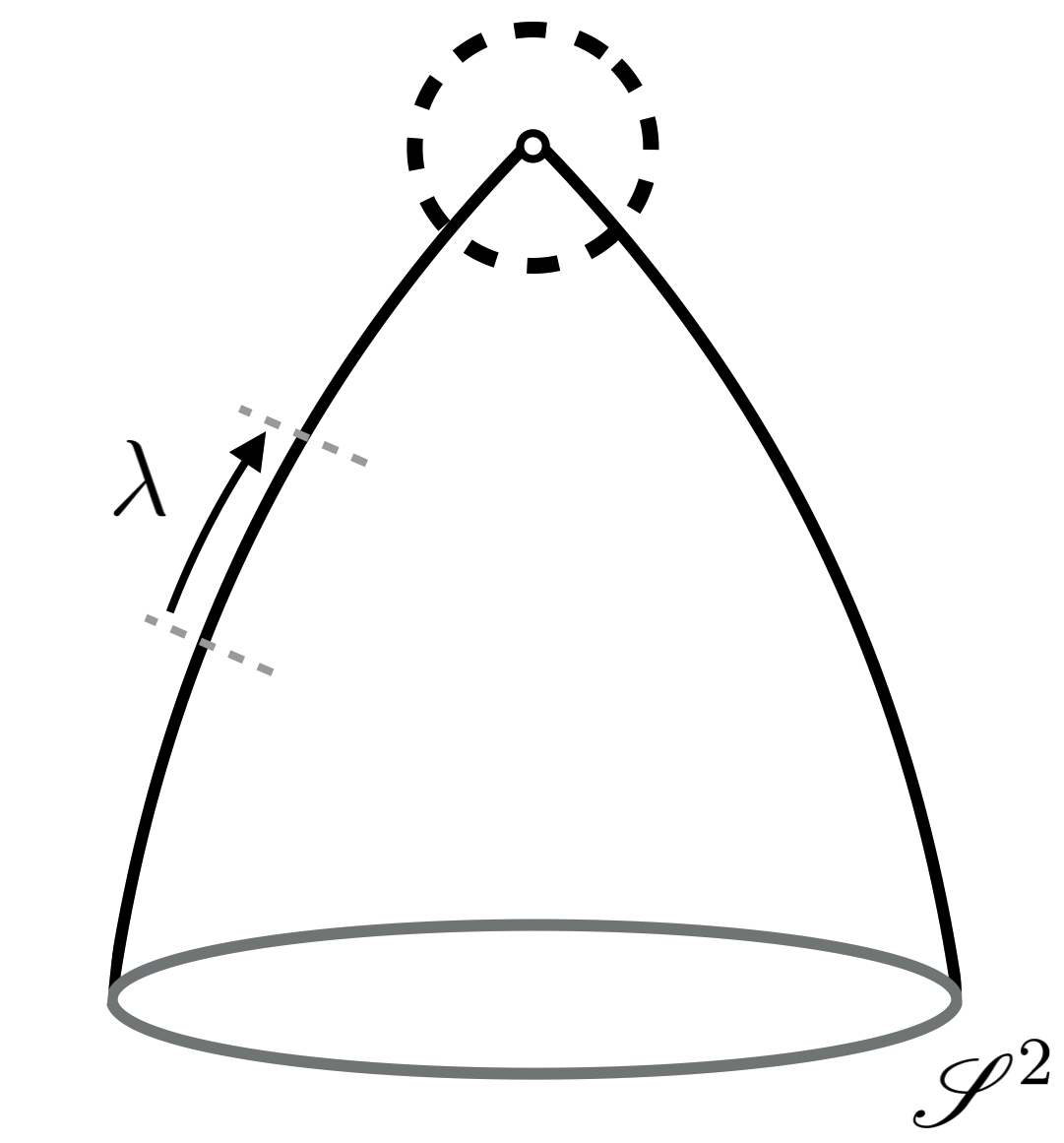}}
\bigskip%
\caption{Outgoing spheres of light with origin in a trapped surface $\mathscr{S}^2$~\cite{Hawking1973} (in our spherically symmetric setting, any $\mathbb{S}^2$ inside the black hole) are convergent. The Einstein field equations and the null energy condition for matter fields lead to the development of a focusing point ($r=0$ in spherical symmetry), which is incompatible with (1-3). Geodesic completeness can be salvaged if the dynamics is modified in an open set around the focusing point (the boundary of which is indicated by a dashed circle).}
\label{fig:1}%
\end{center}
\end{figure}%

Additionally, our spacetimes of interest describe the collapse of a regular distribution of matter from a given initial Cauchy surface with topology $\mathbb{R}^3$. The formation of the black hole entails the formation of a trapped region, defined below.

\noindent
\textsl{Beyond Penrose's theorem.--}Penrose showed that the assumptions (1-3) above, together with the Einstein field equations and the null energy condition for matter fields, result in a contradiction with the existence of a trapped region~\cite{Penrose1964}. In a nutshell (see Fig.~\ref{fig:1}), the reason is that the latter additional dynamical constraints imply the existence of a focusing point at a finite distance (using a suitable definition of distance introduced below), which is incompatible with (1-3). Penrose concluded that, under the assumptions of the theorem, black holes cannot be geodesically complete.

Due to the dimensionality and symmetry of $\mathscr{M}$, there are two spheres of light (radial null geodesics) passing  through every point in spacetime, which we call in\-going and outgoing. In spacetimes with weak gravitational fields, these spheres behave as one would expect from these names: the area of the ingoing ones decreases, while the area of the outgoing ones increases. This behavior is parametrized by the expansions along ingoing and outgoing radial null geodesics, $\theta^{(\bm{k})}(\lambda)$ and $\theta^{(\bm{l})}(\lambda)$, respectively, measuring the infinitesimal rate of change of the area at different points of these geodesics (each of them parametrized in terms of an affine parameter $\lambda$, and with tangent null vectors $\bm{k}$ and $\bm{l}$, respectively). Inside a black hole, the trapped region is defined by the condition of both expansions being negative, while $\theta^{(\bm{l})}(\lambda)$ becomes infinitely negative at the focusing point. In fact, a focusing point at a finite affine distance is unequivocally associated with $\theta^{(\bm{l})}(\lambda)$ becoming negative and divergent~\cite{Wald1984}.

\begin{figure}[!h]%
\begin{center}
\vbox{\includegraphics[width=1.0\columnwidth]{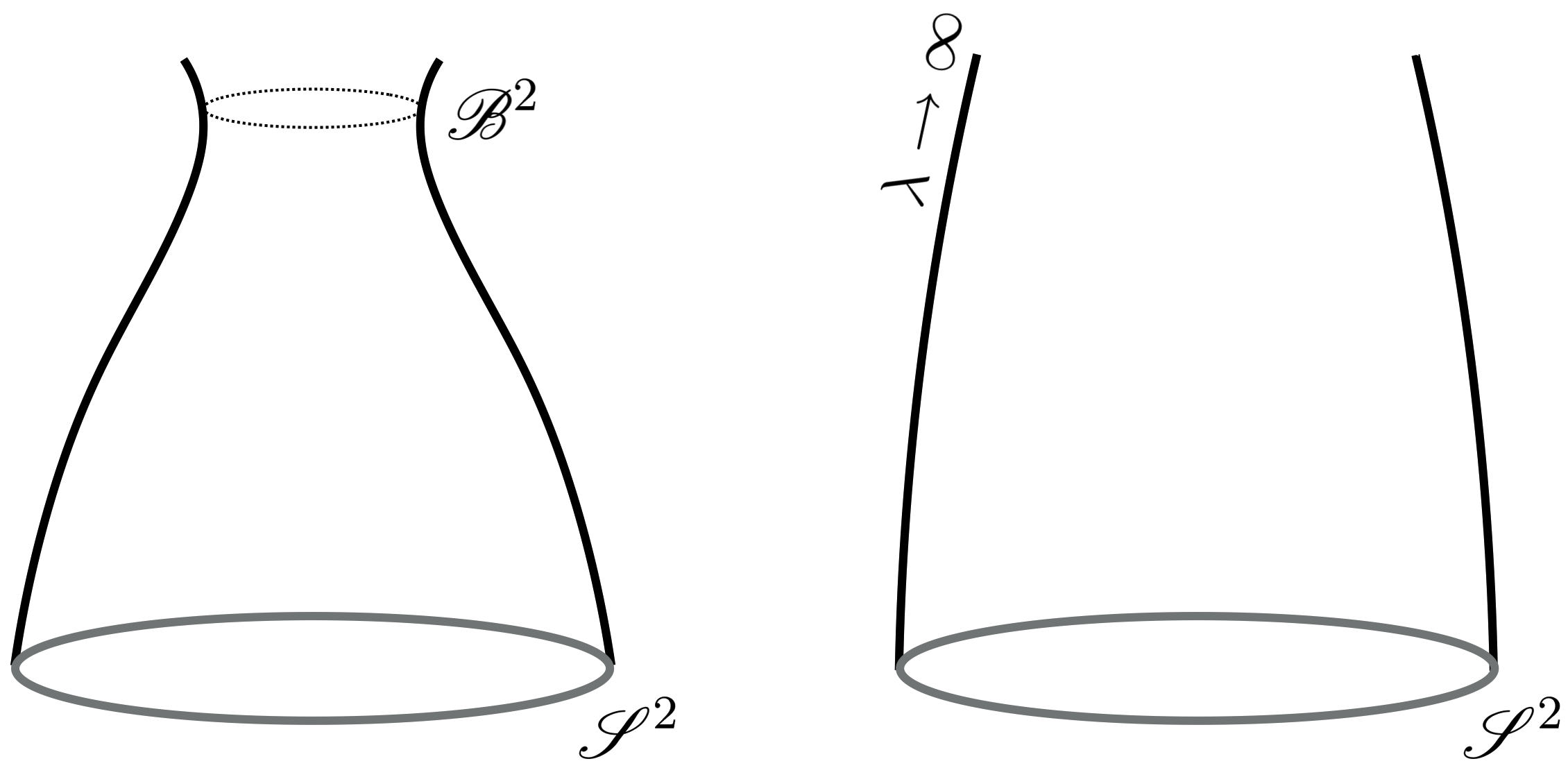}}
\bigskip%
\caption{To guarantee geodesic completeness, either a defocusing point is created at a finite affine distance (thus also creating the 2-surface $\mathscr{B}^2$) or infinite affine distance, or the focusing point is displaced to infinite affine distance. The figure on the right is compatible with both focusing and defocusing points at infinite affine distance.}
\label{fig:2}%
\end{center}
\end{figure}%

Thus, for our purposes, the fundamental insight to be taken from Penrose's theorem is that singularity regularization will either remove the focusing point (thus creating a defocusing point instead), or push the focusing point out to infinite affine distance (see Fig.~\ref{fig:2}). This must be the case for every outgoing radial null geodesic traversing the trapped region.

A defocusing point is defined by the vanishing of $\theta^{(\bm{l})}(\lambda)$, which can occur either at finite or infinite affine distance. Pushing the focusing point to infinite affine distance entails that $\theta^{(\bm{l})}(\lambda)$ remains negative, being perhaps divergent in the limit $\lambda\rightarrow\infty$. It is straightforward to realize that these options include all possible behaviors of outgoing spheres of light in the absence of a focusing point at finite affine distance.

While pushing the focusing point to infinite affine distance allows the spacetime to be geodesically complete, it turns out that this will violate our assumption (4). It is straightforward to show this by studying the regularity (in the limit $r\rightarrow0$) of various curvature invariants (e.g., Ricci and Kretschmann scalars) for the most general spherically symmetric metric. This spacetime can be written without loss of generality as
\begin{align}\label{eq:linel1}
ds^2&=g_{vv}(v,x)dv^2+2g_{vx}(v,x)dvdx+r^2(v,x)d\Omega^2,
\end{align}
where $d\Omega^2$ is the line element on the unit 2-sphere and $0<|g_{vx}|<\infty$ for the metric tensor to be nondegenerate. Indeed, regularity of curvature invariants requires (as a straightforward extension of the argument in~\cite{Carballo-Rubio2018} shows) that, once a trapped region forms, either the function $r(v,x)$ with constant $v$ has a non-vanishing global minimum, or that $g_{vv}(v,x)$ vanishes at least once. 

On the other hand, a direct calculation using the null vector field tangent to outgoing spheres of light, and the corresponding geodesic equation [both can be directly obtained from Eq.~\eqref{eq:linel1}], shows that
\begin{equation}\label{eq:outexp}
\theta^{(\bm{l})}=-\frac{2}{r(v,x)}\frac{g_{vv}}{(g_{vx})^2}\partial_xr(v,x).
\end{equation}
Hence, regularity of curvature invariants implies that $\theta^{(\bm{l})}(\lambda)$ vanishes, either at finite or infinite affine distance.

Our assumptions leave us then with two possibilities. However, these are not completely independent, as one is just the limiting case of the other. Hence, we can focus in the following on the situation in which there is a defocusing point at finite affine distance. This must be the case for all outgoing spheres of light inside the black hole or, equivalently, the (compact) trapped region $\mathscr{T}^4$. These outgoing spheres of light will find in their future a hypersurface $\mathscr{D}^3$ in which their expansion vanishes, which will be part of the boundary of $\mathscr{T}^4$. To determine the properties of $\mathscr{D}^3$ we just need to take into account that there are two independent ways in which the expansion $\theta^{(\bm{l})}$ in Eq.~\eqref{eq:outexp} can vanish: either $g_{vv}(v,x)$ or the derivative $\partial_x r(v,x)$ must vanish. These two independent ways translate into very different behaviors of ingoing spheres of light, as it can be read from their expansion:
\begin{equation}\label{eq:inexp}
\theta^{(\bm{k})}=-\frac{2}{r(v,x)}\partial_xr(v,x).
\end{equation}
Thus, the behavior of both outgoing and ingoing spheres of light traveling across the trapped region $\mathscr{T}^4$ is tightly constrained under the minimal assumptions (1--4).

For the sake of completeness and reproducibility, let us write explicitly the vector fields that we have used above:
\begin{align}
&\bm{l}=\frac{1}{g_{vx}}\partial_v-\frac{g_{vv}}{2(g_{vx})^2}\partial_x,\nonumber\\
&\bm{k}=-\partial_x.
\end{align}
These null vector fields satisfy the usual normalization condition, namely $\bm{l}\cdot\bm{k}=-1$.

\begin{figure}[!h]%
\begin{center}
\vbox{\includegraphics[width=0.5\columnwidth]{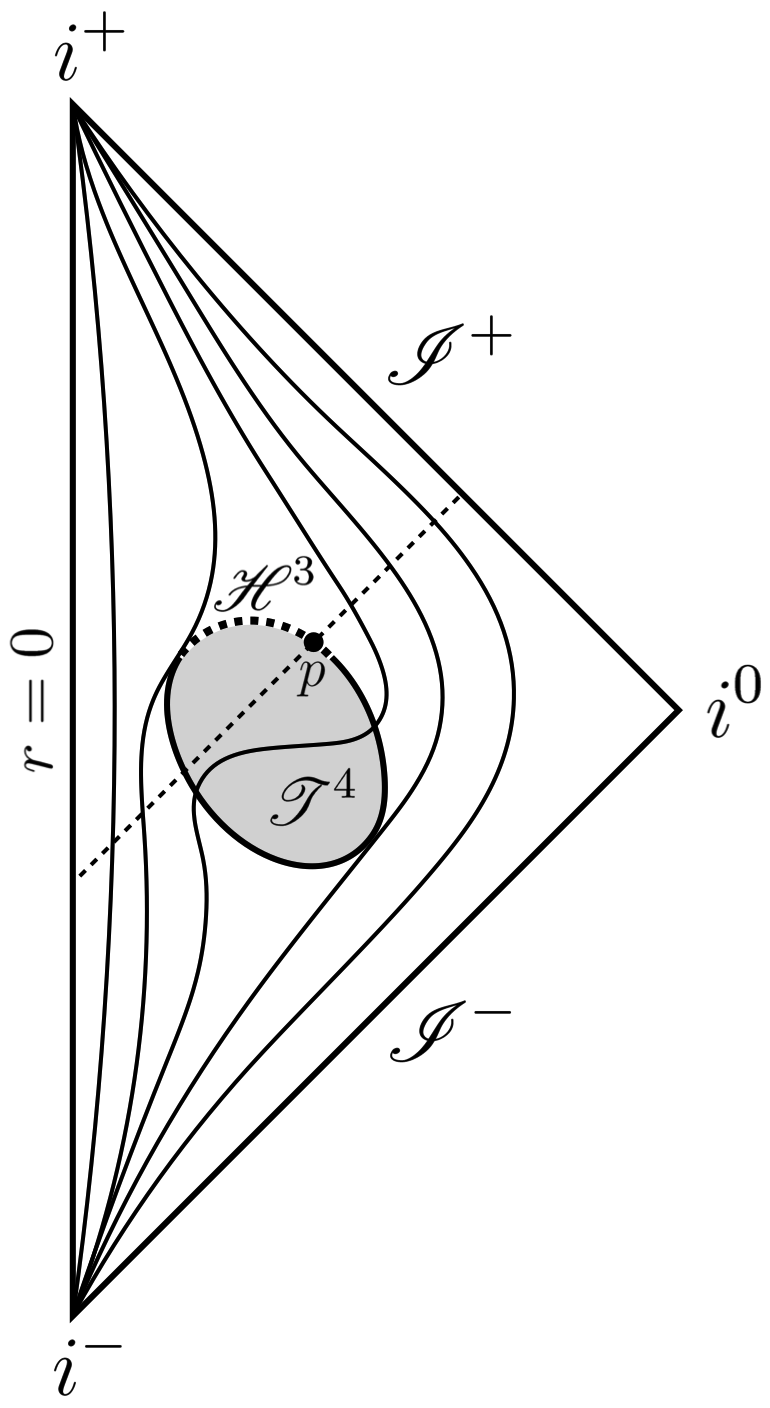}}
\vspace{-0.2cm}
\caption{Penrose diagram of a black hole with evanescent horizons. The dashed line is an outgoing radial null geodesic, while the remaining curved lines mark the hypersurfaces of constant radius $r$.}
\label{fig:3}%
\end{center}
\end{figure}

\noindent
\textsl{Evanescent horizons.--}The first case we analyze is the one in which $\mathscr{D}^3\subset \partial\mathscr{T}^4$ is characterized by $g_{vv}(v,x)=0$, so that $\left.\theta^{(\bm{l})}\right|_{\mathscr{D}^3}=0$ and $\left.\theta^{(\bm{k})}\right|_{\mathscr{D}^3}<0$. The Penrose diagram of the simplest realization of this class of geometries, which seems to have been first analyzed in the Roman--Bergmann article~\cite{Roman1983}, displays a simply connected $\mathscr{T}^4$ as depicted in Fig.~\ref{fig:3}; the most general situation would have several disconnected trapped regions.

The boundary of the trapped region, $\partial\mathscr{T}^4$, is intersected twice by both outgoing and ingoing radial null geodesics. There is a subset of this boundary, $\mathscr{H}^3\subset\mathscr{D}^3\subset\partial\mathscr{T}^4$, containing the second intersection points with ingoing spheres, which is a future inner trapping horizon~\cite{Hayward1993} in which, by definition, $\theta^{(\bm{l})}$ vanishes and its derivative along $\bm{k}$ is positive. Indeed, using Eqs.~\eqref{eq:linel1} and~\eqref{eq:outexp} it is straightforward to show that \mbox{$\left.\mathcal{L}_{\bm{k}}\theta^{(\bm{l})}\right|_{\mathscr{H}^3}=(\partial_xg_{vv})/r(g_{vx})^2>0$}.

\vspace{-0.5cm}

As a consequence, the surface gravity $\kappa_-(v)$ of $\mathscr{H}^3$ is negative, so that $\mathscr{H}^3$ acts as an exponential attractor of nearby geodesics. Extending the analysis in~\cite{Carballo-Rubio2018}, if we integrate the differential equation for outgoing radial null geodesics obtained from Eq.~\eqref{eq:linel1} around its intersection $p\in\mathscr{M}$ with $\mathscr{H}^3$ during an interval $\Delta v$ in which the radius of $p$ remains roughly constant, the coordinate distance $\Delta x(\Delta v)$ to this intersection satisfies
\begin{equation}\label{eq:expg}
\Delta x(\Delta v)\propto \exp\left(-\Delta v|\overline{\kappa}_-|\right),
\end{equation}
where $\overline{\kappa}_-$ is the average value of the surface gravity \mbox{$\kappa_-(v)=\left.-\partial_xg_{vv}/2g_{vx}\right|_{p}<0$} on $\Delta v$. This is a clear indicator of unstable behavior, with instability time scale $1/|\bar{\kappa}_-|$ (which becomes $1/|\kappa_-|$ under the adiabatic condition $|\partial_v\kappa_-(v)|\ll \kappa_-(v)^2$~\cite{Barcelo2010,Barcelo2010b}).
Unless this quantity is fine-tuned, its natural value is roughly proportional to the value of the radial coordinate at $p$ divided by the speed of light $c$ which, in scenarios motivated by quantum gravity, yields the Planck time (see~\cite{Carballo-Rubio2018} for case-by-case calculations). 

The huge discrepancy between this time scale and the naive evaporation time scale due to the emission of Hawking radiation guarantees that these adiabatic conditions would be satisfied (to make things worse, the corrected evaporation time scale that takes into account singularity regularization is actually infinite~\cite{Carballo-Rubio2018}). This leaves no doubt that either our assumptions (1-4) break down at some point during the evaporation process, or there must exist other dynamical processes that become more important than Hawking evaporation and cause the disappearance of $\mathscr{T}^4$ in shorter time scales. 

The instability time scale being typically Planckian points towards a lifetime proportional to the classical collapse time, and therefore linear in the mass, which naively seems to be in flagrant contradiction with astronomical observations. However, that the trapped region disappears (perhaps temporarily) is not equivalent to the complete dispersion of the matter and energy forming the black hole, the distribution of which may remain compact enough to pass current tests~\cite{Carballo-Rubio2018b}. In this scenario, there should be a cycle of formation and disappearance of trapped regions, which from an observational perspective would resemble pulsations of the core of the gravitational potential, induced by multiple bounces of the collapsing matter~\cite{Barcelo2014a,Barcelo2014b,Barcelo2015}. Each of these cycles can be described by a diagram such as the one in Fig.~\ref{fig:3}, but with an additional anti-trapped region in which $\theta^{(\bm{k})}$ changes sign (which makes the diagram symmetric in time). 

These pulsations should yield characteristic phenomenological predictions, and may decay towards a figure of equilibrium~\cite{Barcelo2016}; for instance, it has been proposed ~\cite{Barcelo:2007} that vacuum polarization effects can lead to horizonless structures described by solutions of the semiclassical Einstein field equations~\cite{Carballo-Rubio2017}. Overall, it is clear that, while it is possible to escape the instability issue, this must entail a radical change of our perspective on the nature and evolution of black holes.

\noindent
\textsl{Hidden wormholes.--}The alternative is that $\mathscr{D}^3\subset \partial\mathscr{T}^4$ is characterized by $\partial_xr(v,x)=0$, which implies that $\left.\theta^{(\bm{l})}\right|_{\mathscr{D}^3}=\left.\theta^{(\bm{k})}\right|_{\mathscr{D}^3}=0$ and that there is a region in spacetime where the area of the spheres $\mathbb{S}^2$ is bounded from below, with the minimum contained in $\mathscr{D}^3$. As a consequence of the transition from initial data in which the function $r(v,x)$ had, for $v$ fixed, no global minimum, to a situation in which a global minimum is generated (to guarantee that curvature invariants remain finite in the presence of a trapped region), the spacetime manifold develops a boundary $\partial\mathscr{M}$.

If the causal future of $\partial\mathscr{M}$, $J^+(\partial\mathscr{M})$, is not empty (we shall consider a representative of the empty case below), the area of spheres in this region is bounded, implying that the topology of spacelike hypersurfaces crossing $J^+(\partial\mathscr{M})$ must be $\mathbb{R}\times \mathbb{S}^2$. Hence, realizing this situation dynamically in stellar collapse would require changing the topology of Cauchy surfaces from $\mathbb{R}^3$ to $\mathbb{R}\times\mathbb{S}^2$. This is not compatible with global hyperbolicity~\cite{Hawking1973,Geroch1970,Bernal2003}, which means that these spacetimes cannot accommodate the initial conditions for a realistic collapse \cite{Dafermos2008}. This is entirely analogous to the fact that a realistic collapse cannot create the Einstein--Rosen bridge occurring in the maximally extended Schwarzschild solution \cite{Geroch1971,Dafermos2008}, the Einstein--Rosen bridge has to be imposed \emph{ab initio}.

Let us mention that this kind of geometric structure has been described in situations in which it leads to a bounce into another universe~(see \cite{Ashtekar2018,Ashtekar2018b} and~\cite{Simpson2018,Simpson2019}). However, the spacetimes discussed in these papers are not globally hyperbolic, but display \emph{partial} Cauchy surfaces~\cite{Hawking1973} with topology $\mathbb{R}\times\mathbb{S}^2$.

\noindent
\textsl{To infinity and beyond--}For the sake of completeness, let us briefly discuss the situations in which the defocusing point lies at infinite affine distance. 
The corresponding geometries are limiting cases of the ones already discussed. When taking the limit of infinite affine distance in the ``evanescent horizons" scenario, one gets instead ``everlasting horizons": the black hole never quite disappears.  In the same limit, the ``hidden wormhole" is reduced by half, its throat being pushed to infinite affine distance so that $\mathscr{D}^3 \subset \partial\mathscr{M}$. Topology change is not necessary if $\partial \mathscr{M}$ is chosen so that $J^+(\partial\mathscr{M})$ is empty. 

However, the scenarios above can be made compatible with quantum field theory only if Hawking radiation switches off asymptotically. A vanishing Hawking temperature requires a vanishing surface gravity, implying the asymptotic onset of an extremal configuration, which in the ``everlasting horizons" scenario requires the asymptotic merger of inner and outer horizons, while it is incompatible with the two classes of ``hidden wormholes" without the breakdown of standard quantum field theory at macroscopic scales. 

\noindent
\textsl{Conclusions--}Our assumptions (1-4) lead to a remarkably limited set of possibilities for non-singular black hole spacetimes. A technical limitation of our analysis is that these assumptions disregard the possibility that the description in terms of pseudo-Riemannian manifolds may break down due to fluctuations of the spacetime geometry. It has been suggested (e.g.,~\cite{Ashtekar2005,Perez2014,Visser1997}) that this will indeed be the case, which may open new possibilities for singularity regularization. On the other hand, it is also true that advances in loop quantum cosmology show that a description in terms of an effective metric provides a good approximation in many situations~\cite{Kaminski2010,Ashtekar2011,Ashtekar2015} so that, for a large sector of initial conditions for the quantum state of the system, the description in terms of differentiable manifolds remains meaningful throughout dynamical evolution. We think that there is no question that effective descriptions in terms of differentiable manifolds satisfying (1-4) can provide important insights on the physics at play.

Having said that, we cannot discard the possibility that genuine quantum gravity effects may cure the self-consistency issues raised here. It might be the case that the apparent fine-tuning of the surface gravity of the inner horizon, necessary to tame its unstable behaviour in the ``evanescent horizons'' or ``everlasting horizons'' scenarios, turns out to be dictated by the underlying ultraviolet completion. Or, it might be the case that quantum fluctuations disrupt Hawking radiation towards the end of the lifetime of ``hidden wormholes".

However, the final point we want to make in this Letter is that there is no middle ground: either future analyses realize these alternative possibilities, or we will face a complete rethinking of the theoretical concept of black hole, along with significant consequences for future astronomical observations.

\acknowledgments

\noindent
\textsl{Acknowledgments--}The authors thank Carlos Barcel{\'o} and Javier Olmedo for helpful discussions. MV was supported by the Marsden Fund, which is administered by the Royal Society of New Zealand. MV would like to thank SISSA and INFN (Trieste) for hospitality during the early phase of this work.

\bibliography{refs}	

\end{document}